\documentstyle[preprint,prl,aps,epsf]{revtex}
\begin{document}
\draft
\preprint{}
\title{Properties of the random field Ising model in a transverse
magnetic field}
\author{T. Senthil}
\address{Department of Physics, Yale University,
P.O. Box 208120,
New Haven, CT-06520-8120\\
and\\
Institute for Theoretical Physics, 
University of California,
Santa Barbara, CA 93106-4030 \thanks{Address since September 1997}\ }
\date{\today}
\maketitle
 
\begin{abstract}
 We consider the effect of a random longitudinal field on the Ising
model in a transverse magnetic field. For spatial dimension $d > 2$,
there is at low strength of randomness and transverse field, a phase
with true long range order which is destroyed at higher values of the
randomness or transverse field. The properties of the quantum phase transition 
at
zero temperature are controlled by a fixed point with no 
quantum fluctuations. This fixed point also controls the {\em classical}
finite temperature phase transition in this model. Many critical 
properties of the quantum transition are therefore identical to those
of the classical transition. In particular, we argue that the dynamical 
scaling is activated, {\it i.e}, the logarithm of the diverging time scale
rises as a power of the diverging length scale.

\end{abstract}
%\vspace{0.25cm}
 

\narrowtext
%\newpage

A number of recent theoretical \cite{Fisher,Miller,Sachdev,Bhatt,Potts,percol} 
and experimental \cite{Rosen} works have studied the effects of
randomness on simple quantum statistical models. These studies have
been motivated by the need to understand in a simple context the interplay
of effects related to strong randomness, interactions, and quantum
fluctuations. In this paper, we study the effect of a random longitudinal
field applied to the transverse field Ising model. The effects of random
longitudinal fields on {\em classical} Ising models have been 
studied extensively\cite{review}, and are partially well understood.
In contrast, very little is known about the corresponding
quantum problem in realistic dimensions. Most previous studies
of this system have been limited to mean field theory (expected to
be valid above $6$ spatial dimensions; see below) or to
an expansion in $\epsilon = 6 -d$. However, as is well known 
from the classical problem, these results are not expected to
directly be of much relevance to physical systems in
finite dimensions much smaller than $6$. Here we will
provide  a general scaling theory applicable in any dimension.

The model in consideration is defined by the Hamiltonian
\begin{equation}
\label{qmodel}
{\cal H} = -J \sum_{i,j} \sigma_i^z \sigma_j^z - h\sum_i\sigma_i^x
            - \sum_i H_i \sigma_i^z
\end{equation}
For simplicity, we have assumed that $J$ and $h$ are non-random,
though our results should apply also for weakly random $J$ and $h$.
$H_i$ is assumed to be random with zero mean and variance $\Delta ^2$. 
The $\vec \sigma_i$ are
Pauli spin matrices. We note that, as usual, this quantum model in 
$d$ dimensions at $T = 0$
is equivalent to a {\em classical} Ising model in $d + 1$  
dimensions in a random field
with the randomness correlated along one direction.

It will often be more convenient to consider a coarse-grained
continuum field theoretic version of the Hamiltonian Eqn.~\ref{qmodel}.
The continuum action is readily written down as:
\begin{equation}
\label{action}
S = \int d\tau \int d^d x \frac{1}{t}\left[(\frac{\partial \phi}
{\partial \tau})^2 + (\nabla \phi)^2 + r\phi^2 +u\phi^4  - 
H(x)\phi(x,\tau)\right]
\end{equation}
where $H(x)$ is the (coarse-grained) random field and is taken
to be Gaussian distributed with mean $0$ and variance $\Delta^2$.
For future convenience, we have introduced a factor $t$ as an
overall scale for the action. 
 
\begin{figure}
\epsfxsize=8in
\centerline{\epsffile{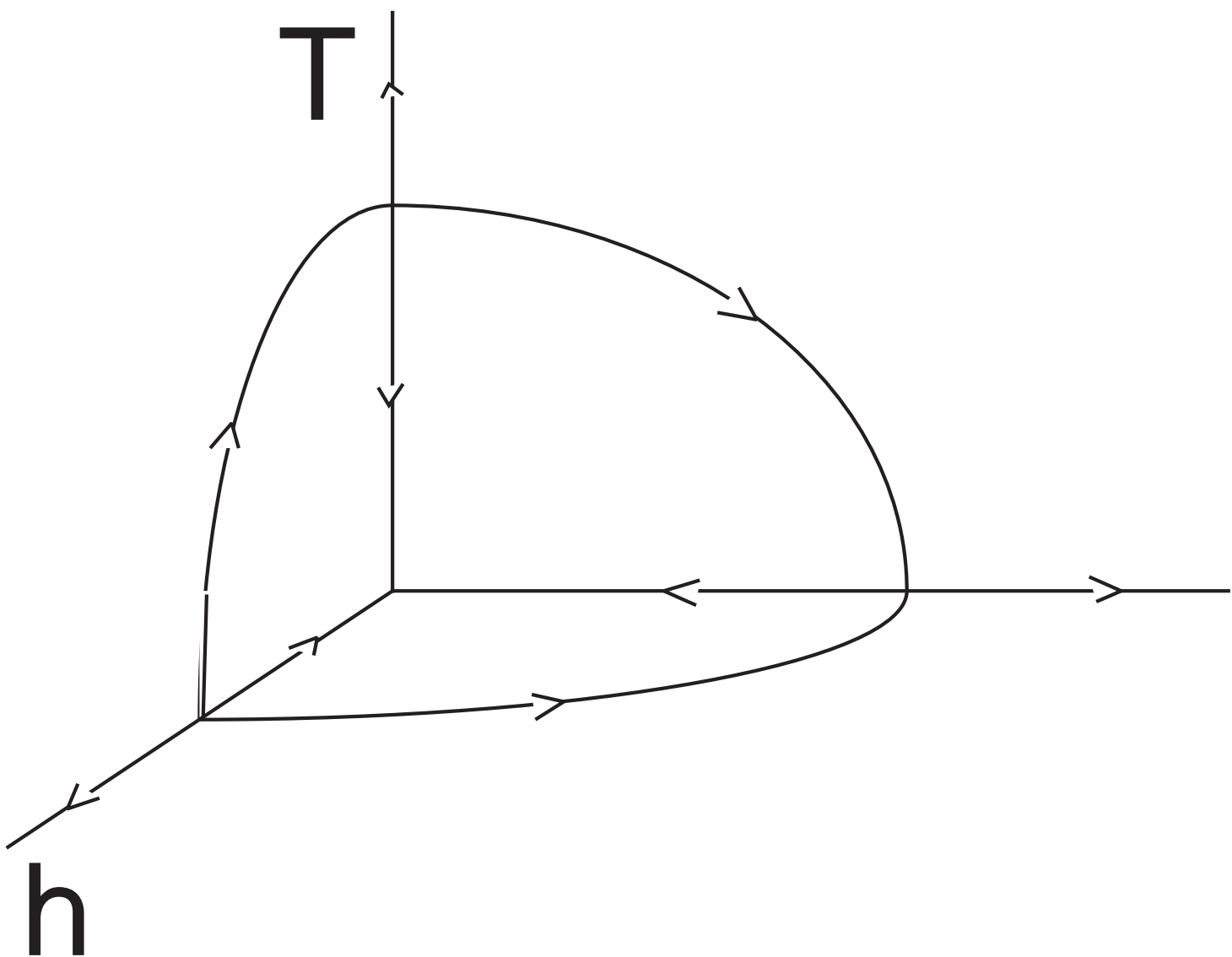}}
\vspace{0.5in}
\caption{Schematic phase diagram and renormalization group flows
for the transverse field Ising model in a random longitudinal field
for $d > 2$ in the temperature($T$), transverse field($h$), strength
of randomness($\Delta$) space. There is a transition from a ferromagnet
to a paramagnet as any of these three parameters is increased.}
\vspace{0.5in}  
\label{rfield}
\end{figure}

 Later in this paper, we use the standard Imry-Ma argument
to show that the ordered phase is unstable to 
any weak random field for spatial dimension $d \leq 2$.
In Fig~\ref{rfield}, we show the expected phase diagram
for $d > 2$ in the temperature ($T$), $h$, $\Delta$ space.
In what follows, we will concentrate primarily on the $T = 0$
transition from the ordered phase to the paramagnetic phase.
Our central claim is that the quantum transition is controlled
by a fixed point with no quantum fluctuations in any dimension
($> 2$). However quantum fluctuations are ``dangerously irrelevant"
at this fixed point, and need to be included to get 
the correct critical behaviour of many physical observables.
Such fluctuationless fixed points have arisen in several other
recent studies of quantum transitions in random 
systems\cite{Fisher,sro,Potts,percol,Belitz}. For the  model in
consideration here,
this claim is completely analogous to the corresponding claim on the role
of thermal fluctuations at the finite temperature phase transition
in {\em classical} random field Ising magnets\cite{review}.
Indeed, we claim that the {\em same fixed point} controls both the
classical (finite $T$) and quantum ($T = 0$) transitions in this
model. This enables us to relate many of the critical properties of the
quantum transition to those of the well-studied (but only
partially understood) classical transition. 
In particular, we suggest that the dynamical 
scaling is activated, {\it i.e}, the logarithm of the diverging time scale
rises as a power of the diverging length scale.

\section{Weak randomness}
 We first consider the effects of weak
random fields on the properties of the pure system. 
These are quite innocuous deep in the paramagnetic phase
when the transverse field wins over both the randomness
and the exchange interaction. So we will turn
immediately to the ordered phase and the critical point.

It is easy to see using the Imry-Ma\cite{Imry,note,Imbrie} argument that 
for dimension $d > 2$, the ordered phase of the quantum model
is stable to a  weakly random $H_i$. Imagine that in the 
absence of randomness, the system is deep in the ordered 
phase. The stability to a weak random field is 
determined by balancing the energy cost to form 
large-sized domains with the typical energy gain because of
the random $H_i$. Deep in the ordered phase, the energy cost of a 
domain $\sim L^{d - 1}$ while the typical energy gain due to the 
random field $\sim L^{d/2}$ exactly as in the classical 
problem. Thus for $d > 2$, for weak enough random fields,
it is not favourable to form large-sized domains and the system
stays ordered. We expect of course that strong randomness
will destroy the ordered phase in any dimension. Precisely 
in $d = 2$, as in the classical problem, the system is 
marginally unstable and there is no ordered phase.

Now, assume that we are at the critical point of the pure system
in $d>2$. When are weak random fields a relevant perturbation
at this critical fixed point? This can be answered as follows.
Consider the continuum action Eqn.~\ref{action}. Averaging
over the disorder using replicas gives the term
$\frac{\Delta^2}{2}\int d^d x \int d\tau_1 d\tau_2 
\sum_{a,b}\phi_a(x,\tau_1)\phi_b(x,\tau_2)$. Now under the
renormaliation group transformation appropriate to the 
critical fixed point of the pure system $x \rightarrow x' = \frac{x}{s}$,
$\tau \rightarrow \tau' = \frac{\tau}{s^z}$, and
$\phi \rightarrow \phi' = \phi s^{\frac{d+z-2+\eta}{2}}$.
This gives $\Delta^{2\prime} = \Delta^2 s^{z+2-\eta}$. Clearly
then $\Delta$ is always relevant.

\section{The critical fixed point}
In this section, we will present the reasoning behind
our assumption that the fixed point controlling the
transition has no quantum fluctuations\cite{note1}.

For {\em classical} random field magnets, it is believed\cite{review}
that the properties of the finite temperature phase transition
(which exists for $d > 2$) are controlled by a zero temperature
fixed point. The first evidence for this belief came from
perturbative studies\cite{Imry,dimred} of a continuum field theoretic
description of the magnet near the critical point. Order by
order in perturbation theory, it can be shown that the
effects of the fluctuations introduced by the randomness
dominate over the effects of the thermal fluctuations
near the critical point. This is interpreted in
renormalization group language to be a manifestation
of a zero-temperature fixed point with (dangerously)
irrelevant thermal fluctuations. Though many of the 
predictions of this perturbative analysis are 
believed not to be correct for realistic dimensions ($d=2$,$3$),
there is a general consensus that the critical
fixed point is fluctuationless in all dimensions
where there is a transition. This is further
supported by heuristic renormalization group
arguments valid in $d = 2+\epsilon$\cite{Bray1},
numerical real space renormalization calculations\cite{Newman},
and general agreement of some of the phenomenology suggested
by such fluctuation-less fixed points with experiments in
$d=3$\cite{Belanger}.

For the quantum random field magnets that we are considering here,
a similar perturbative analysis, valid in high dimension, 
of a continuum field theory expected to describe the correct $T=0$
critical behaviour was undertaken several years ago by two
groups\cite{Aharony,Cardy}(See also Section~\ref{epsilon}). 
Again it was found that order by 
order in perturbation theory, the effects of fluctuations
introduced by the randomness dominated the effects of 
quantum fluctuations. We take this to be strong evidence
that the fixed point is fluctuationless for every dimension
($d>2$) for the quantum problem as well.

Now consider the finite temperature phase transition in the
quantum model which occurs when the ground state is ordered.
This transition is, of course, in the same universality
class as that in the {\em classical} models discussed
earlier. 
It is clear that this is controlled by the same fixed point
that controls the quantum transition. Thus, we have the 
unusual situation that the same fixed point controls
both the $T=0$ and $T \neq 0$ transitions. The phase 
diagram and RG flows (for $d>2$) are shown in Figure~\ref{rfield}.

\section{General Scaling Hypothesis for $2 < d < 6$}
A number of results follow from the claim that the same fixed
point controls both the classical and quantum transitions.
First, on approaching the transition from the ordered phase, 
the magnetization vanishes with an exponent $\beta_Q$  which is the 
same as for the classical transition. Next consider
correlation functions. As in the classical problem,
there are
two different correlation functions that can be defined:
The  correlation function
\begin{equation}
C(x,\tau;x',\tau') = \overline{<\sigma_z (x,\tau)> <\sigma_z (x',\tau')>}
-\overline{<\sigma_z(x,\tau)>}~\overline{<\sigma_z(x',\tau')>}             
\end{equation}
where the angular brackets denote averaging over quantum
fluctuations and the overline denotes averaging over the 
randomness. As the randomness is independant of time, the
system is translationally invariant in time in every sample.
Therefore $C(x,\tau;x',\tau')$ is independant of $\tau, \tau'$
and we will refer to it as $C(x,x')$ from now on. After averaging
over the disorder, spatial translational invariance is also
restored in the correlation function, and so $C(x,x') = C(x-x')$.
This satisfies (near the transition) 
\begin{equation}
C(x) \sim \frac{1}{x^{d - 2 + \overline \eta_Q}} 
                    {\cal C}\left(\frac{x}{\xi}\right)
\end{equation}
where the correlation length $\xi \sim |h - h_c|^{-\nu_Q}$. The
scaling function ${\cal C}$ and the exponents $\nu_Q$, $\overline \eta_Q$ 
are properties of the fixed point
theory and its relevant perturbations (the deviation from the
critical randomness strength). As these are the same for both the 
quantum and classical transitions, we get the result that the
exponents $\nu_Q$, $\overline \eta_Q$, and the function ${\cal C}$
are identical to their classical counterparts.

The connected correlation function
\begin{equation}
G_c (x,\tau;x',\tau') = \overline{ <T_{\tau}
(\sigma_z (x,\tau) \sigma_z(x',\tau')) > - 
              < \sigma_z(x,\tau) > < \sigma_z(x',\tau') >}
\end{equation}
where $T_{\tau}$ is the time-ordering symbol.
Clearly $G_c(x,\tau;x',\tau') = G_c(x - x',\tau-\tau')$. 

Note that this correlation function vanishes at the fixed point (as there
are no quantum fluctuations at the fixed point). Thus to obtain its critical
behaviour, we need to keep the irrelevant quantum fluctuations. Similarly
for the corresponding classical problem, it is necessary to
keep the irrelevant thermal fluctuations. As these two irrelevant
perturbations may have quite different effects, in general, we {\em do not}
expect $G_c$ to scale identically to the  classical space and time dependant
correlation function.

However it is possible to argue that the {\em static} correlation
function of the quantum problem scales identically to the {\em equal-time}
correlation function of the classical problem. This was first noticed
within the $\epsilon$ expansion by Boyanovsky and Cardy\cite{Cardy}, but as we
show below is more generally valid (though other results of the $\epsilon$
expansion are not). Consider the susceptibility\cite{note2}of the model to an
external spatially varying static magnetic field $H_{ext}(x)$. 
Clearly for both the
quantum and classical problems, the scaling of this susceptibility
near the transition is just determined by minimizing the classical
fixed-point Hamiltonian in the presence of the external field. Hence
the static (non-local) susceptibilility scales identically for the 
classical and quantum problems. Now at any finite $T \neq 0$ in the classical
problem, the fluctuation-dissipation theorem implies that the equal-time
correlation function is proportional to the susceptibility. Similarly,
the static correlation function of the quantum problem is also 
proportional to its susceptibility. Thus these two correlation functions
scale identically. We then have the result
\begin{equation}
\int d\tau G_c(x,\tau) \sim \frac{1}{x^{d-2+\eta}}
           \tilde{{\cal G}}\left(\frac{x}{\xi}\right)
\end{equation}
with $\eta$ and the scaling function $\tilde{{\cal G}}$ being identical
to those for the connected equal-time correlation functions of the
classical problem. The equal-time correlation function in the
quantum problem will however scale differently.

We now turn to dynamical correlations. For the classical transition,
Villain and Fisher\cite{Villain} have presented arguments to
show that the dynamical scaling is unconventional, with the 
logarithm of the characteristic relaxation times scaling as a
power of the length scales. Similar unconventional dynamic
scaling also occurs in the vicinity of the random {\em quantum}
transitions studied  recently\cite{Fisher,Potts,percol}, all of which
are controlled by fluctuation-less fixed points. We suggest below
that the dynamic scaling is activated at this quantum transition 
as well.

The argument for the dynamic scaling closely
follows that for the classical case. Consider a block of the 
system of size $\sim \xi$ near the critical system. Ignoring
quantum fluctuations, the energy landscape as a function of the total
magnetization of the block has been 
argued to scale as $\xi^\theta$\cite{Villain}.
Most blocks would thus have a single deep global minimum at some non-zero value 
of the magnetization. The energy of this minimum will differ from 
those of other local mimima by amounts $\sim \xi^\theta$. The
barriers separating these different local mimima also scale as
$\xi^\theta$. Effects of quantum fluctuations on such blocks should be rather
small, and do not contribute significantly to the dynamics. However, there
would be some rare blocks in which there are two minima with an energy 
separation which is nearly zero. 
The dynamics at long time scales will be dominated
by quantum tunnelling between such minima in these rare blocks. As the 
barrier between these minima rises as a power of $\xi$,  it is natural that
the tunneling time $t_Q \sim \exp(c\xi^\psi)$ with $\psi$ a new exponent.
(This quantum tunneling will be important so long as the energy difference
between the two minima is roughly less than $\hbar/t_Q$).
Thus the dynamics is activated like in the other random quantum transitions 
studied in Refs~\cite{Fisher,Potts,percol}. 

This therefore motivates the following scaling form for the imaginary
part of the $q,\omega$ dependant susceptibility (which is the spectral
density for the connected correlator introduced above):
\begin{equation}
\chi''(q,\omega) = \xi^{\kappa}f(q\xi, \frac{\ln{\frac{1}{\omega t_0}}}
{\xi^{\psi}})
\end{equation}
where $t_0$ is a microscopic time scale. The exponent $\kappa$ will be related
to other exponents below. The real part of the susceptibility may
be obtained by the Kramers-Kronig relation:
\begin{displaymath}
\chi'(q,\omega) = \int \frac{d\omega'}{\pi} 
\frac{\chi''(q, \omega')}{\omega' -\omega}
\end{displaymath}
Following Pytte and Imry\cite{Pytte}, we write 
$y = \frac{\ln{\frac{1}{\omega' t_0}}}
{\xi^{\psi}}$ and $z = \frac{\ln{\frac{1}{\omega t_0}}}
{\xi^{\psi}}$. Then
\begin{displaymath}
\chi'(q, \omega) = \xi^{\kappa + \psi} \int_{-\infty}^{+\infty} \frac{dy}{\pi} 
f(q\xi,y)
\left[ \frac{1}{e^{\xi^{\psi}(y-z)}+1} - \frac{1}{e^{\xi^{\psi}(y-z)}-1}\right]
\end{displaymath}
In the scaling limit when $\xi \rightarrow \infty$, we may approximate
\begin{eqnarray*}
\frac{1}{e^{\xi^{\psi}(y-z)}+1} & \approx & \theta(z-y) \\
\frac{1}{e^{\xi^{\psi}(y-z)}-1} & \approx & -\theta(z-y)
\end{eqnarray*}
Thus
\begin{displaymath}
\chi'(q,z) \sim \frac{2}{\pi}\xi^{\kappa+\psi}\int_{-\infty}^{z} dy f(q\xi,y)
\end{displaymath}
Thus we get the following scaling form for the real part:
\begin{equation}
\chi'(q,\omega) \sim \xi^{\kappa + \psi} \tilde f (q\xi, 
\frac{\ln{\frac{1}{\omega t_0}}}
{\xi^{\psi}})
\end{equation}
with $\tilde f(x,z) = \frac{2}{\pi}\int_{-\infty}^{z} dy f(x,y)$.
In particular, the {\em static} susceptibility
\begin{displaymath} 
\chi'(q,0) \sim \xi^{\kappa +\psi}\tilde f(q\xi,-\infty)
\end{displaymath}
Thus we identify $\kappa = 2-\eta-\psi$.
Note that similar scaling forms apply for the dynamics near the 
classical transition as well, although with a different value for 
$\psi$ and a different scaling function $f(x,y)$. Nevertheless, as
we argued earlier the function $\tilde f (x,-\infty)$ should be the same
for the quantum and classical transitions.

\section{Expansion in $\epsilon = 6-d$}
\label{epsilon}
As shown below, the upper critical dimension of this model 
is $6$. It is natural to try an expansion in powers of
$\epsilon = 6-d$. This was done long back by
Aharony, Gefen and Shapir\cite{Aharony} and by Boyanovsky and Cardy\cite{Cardy}.
Here,
we will review their results and discuss them in the context
of the general scaling hypotheses of the previous section.

The $\epsilon$ expansion is done using the continuum action Eqn.~\ref{action}.
It is instructive to set up the renormalization group so that $t$ is
allowed to flow while keeping the strength of the randomness fixed.
 In particular, the results of Ref.~\cite{Aharony,Cardy}
show that $t$ flows to zero at the critical fixed point. When $t=0$,
the partition function is determined by the particular configuration
$\phi(x,\tau)$ that minimizes the action. Clearly the minimum action
configuration is static. Thus solving the fixed point theory simply 
corresponds to finding the static configuraion $\phi_{stat}(x)$ that 
minimizes the potential energy terms in the action. Thus the fixed point theory
is entirely classical.

First consider the Gaussian theory with $u=0$.
The correlation functions $C(k,\omega) = 
\overline{\langle \phi(k,\omega)\rangle
\langle \phi(-k,-\omega)\rangle}$ and $G(k,\omega) = \overline{ \langle 
\phi(k,\omega) \phi(-k,-\omega)\rangle - \langle \phi(k,\omega)\rangle
\langle \phi(-k,-\omega)\rangle}$ can be easily calculated. The result
is:
\begin{eqnarray*}
C(k,\omega) & = & \frac{\Delta^2 \delta(\omega)}{k^2 + r} \\
G(k,\omega) & = & \frac{t}{k^2 + \omega^2 + r}
\end{eqnarray*}
The critical point is at $r=0$. The correlation functions
introduced in the previous section are seen to scale as
\begin{eqnarray}
C(x) & \sim & \frac{1}{x^{d-4}}{\cal C}(\frac{x}{\xi}) \\
G_c(x,\tau) & \sim & \frac{1}{x^{d-1}}{\cal G}(\frac{x}{\xi})
\end{eqnarray}
with the correlation length $\xi = \frac{1}{\sqrt r}$.
Thus, in the Gaussian theory, we have $\nu_Q = \frac{1}{2}$, 
 $\eta_Q = 0$,  $\overline{\eta_Q} = -2$.
Note that for the classical problem also, the Gaussian exponents are
$\nu = \frac{1}{2}$, 
$\eta =0$,  $\overline{\eta} = -2$. This is a trivial illustration of 
the general point made in the previous section. The dynamic scaling
is however conventional in the Gaussian theory. 

Now consider a momentum shell renormalization group transformation
on the Gaussian action. It is convenient to let $t$ flow and keep
the strength $\Delta$ of the random field fixed. The flow equations
for $t$ and $r$ are readily seen to be
\begin{eqnarray*}
\frac{dt}{dl} & = & -3t \\
\frac{dr}{dl} & = & 2r
\end{eqnarray*}
Thus even in the Gaussian theory, $t$ flows to $0$ at the fixed point
which is hence fluctuationless. The tree-level flow equation for $u$
at the Gaussian fixed point is just obtained by power-counting and is
\begin{displaymath}
\frac{du}{dl}  =  (6-d)u
\end{displaymath}
Thus interaction effects are irrelevant above $6$ spatial dimensions
and the Gaussian theory gives the true critical behaviour. For
$d$ below $6$, it is possible to construct an expansion in
powers of $\epsilon = 6-d$\cite{Aharony}. To leading order,
the one loop RG equations are:
\begin{eqnarray*}
\frac{dt}{dl} & = & -3t \\
\frac{dr}{dl} & = & 2r + \frac{3ut}{2}K_d \Lambda^{d-1}(1-\frac{r}{2\Lambda^2}) + 
6 u \Delta K_d \Lambda^{d-6}(1-\frac{3r}{\Lambda^2}) \\
\frac{du}{dl} & = & (6-d)u -\frac{9u^2t}{4}K_d\Lambda^{d-3}
  -36u^2\Delta K_d\Lambda^{d-6}
\end{eqnarray*}
where $\Lambda$ is the high-momentum cutoff and $K_d =  \frac{S_d}{(2\pi)^d}$.
($S_d$ is the surface area of a unit sphere in $d$ dimensions).
Again $t$ flows to $0$ at the fixed point. Setting $t = 0$ in the remaining
equations it is clear that there is a non-trivial fixed point at 
$r* = -\frac{\epsilon \Lambda^2}{12}, u* = \frac{\epsilon}{36\Delta K_6}$.
The flows can be linearized around this fixed point and give, for instance,
$\nu = \frac{1}{2} + \frac{\epsilon}{12}$ to first order in $\epsilon$.
Note that this is the same as for the pure problem in $3-\epsilon$
dimensions. This is a general feature of the $\epsilon$ expansion
- all the exponents characterizing static critical properties in $d$
dimensions
are the same as the pure problem in $d-3$ dimensions. This result however
is an artifact of the $\epsilon$ expansion and is not true in all
dimensions. 

The fact that $t$ flows to $0$ to this order means that the new
non-Gaussian fixed point is also fluctuationless. In fact the flow
equation for $t$ has been shown to be exact to all orders in 
$\epsilon$\cite{Aharony,Cardy}. Thus at least within the $\epsilon$ expansion,
the fixed point is fluctuationless in any dimension.

It was 
argued by Boyanovsky and Cardy that all the exponents and scaling
functions associated with static critical properties of the quantum transition
were the same as for the classical transition. As we have seen in the
previous section, this result is true quite generally. 
However dynamic properties
(for instance time-dependant correlation functions) were shown to 
scale differently. Within the $\epsilon$ expansion, they found that
the dynamic scaling is conventional at both transitions with 
$z_Q = 1 + (0.0185)\epsilon^2 + (0.0182)\epsilon^3 + o(\epsilon^4)$
and $z_{cl} = 2z_Q$. Thus,
 as in the 
classical case, the $\epsilon$ expansion is qualitatively incorrect 
to describe many aspects of the critical behaviour of the
random field quantum system well below the upper critical
dimension. Nevertheless the $\epsilon$ expansion provides
useful evidence for the claim that the fixed point has no
quantum fluctuations. 

\section{Discussion}
 What we have done in this paper is primarily
to update the theory of the quantum random field models since the pioneering
papers in Ref.\cite{Aharony,Cardy}, to take into account subsequent develoments
in the understanding of the classical problem. Our main assumption,
motivated by the results of Ref.\cite{Aharony,Cardy}, was that the quantum
transition in any dimension is controlled by a fluctuationless fixed point
that also controls the classical finite temperature transition. Some of the 
consequences of this assumption were then examined. The static critical
properties of the quantum transition are {\em identical} to those
of the classical transition. The dynamics (which depends on the irrelevant
quantum/thermal fluctuations) is similar, though not identical. 
In particular, we suggested
that the dynamic scaling is of the activated form, with the length scales
depending logarithmically on time scales. We have however not
provided calculational evidence for this suggestion.
Evidence for the activated dynamic
scaling cannot be obtained from the $\epsilon$ expansion
(or from other perturbative approaches like a $2 + \epsilon$
expansion). Even for the classical transition, the only 
available evidence comes from numerical calculations\cite{Huse}
and agreement with the phenomenology seen in experiments\cite{Belanger}.
Here, we have argued that activated dynamics in the quantum case
is  natural if it occurs in the classical problem. We may 
therefore regard support for activated dynamics in the classical
transition as some sort of support for it happening in the
quantum transition as well.

There are some other important questions that are left open. For most of
the paper, we have focused on the critical point. The paramagnetic
phase should be interesting to study by itself. In particular,
is there a finite gap all the way till the critical point, or is there
a Griffiths-phase with gapless excitations in the vicinity of the
critical point? It is possible to show\cite{Sen_GM} for the quantum
version of a special model introduced by Grinstein and Mukamel\cite{GM} 
in $d=1$ that the spin-autocorelation in imaginary time has
a stretched exponential form $e^{-c\tau^{\frac{1}{3}}}$. This implies
gapless excitations with an essential singularity at zero energy in the
local density of states. For more realistic models however, even in $d=1$,
the autocorrelation presumably decays exponentially. The general situation
is unclear. Similar questions can also be asked about Griffiths effects
for the uniform susceptibility in the paramagnetic phase.

\acknowledgments
I thank D.S. Fisher, N. Read, and in particular S. Sachdev for several 
useful discussions
and encouragement. This research was supported by NSF Grants DMR-96-23181
and PHY94-07194.

\begin{references}
\bibitem{note1}{We are always also assuming that the transition 
remains second order in the presence of the random field}.
\end{references}
\end{document}